%% file: main.tex
\pdfoutput=1
\documentclass[letterpaper,twocolumn,10pt]{article}
\usepackage{usenix2019_v3}

\input{packages.tex}
\input{commands.tex}
\input{space-tricks.tex}

\begin{document}

\title{\sys{}: A Trusted Lease Primitive for Distributed Systems}
\author{Bohdan Trach\textsuperscript{1}, Rasha Faqeh\textsuperscript{1}, Oleksii Oleksenko\textsuperscript{1}, Wojciech Ozga\textsuperscript{1}, Pramod Bhatotia\textsuperscript{2}, Christof Fetzer\textsuperscript{1}\\
  \textsuperscript{1}TU Dresden,  \textsuperscript{2}TU M\"unich}
\maketitle

\input{0_abstract}
\input{1_intro}
\input{2_overview}
\input{3_design}
\input{4_implementation}
\input{4_modeling}
\input{5_evaluation}
\input{7_related}
\input{6_discussion}
\input{8_conclusion}
\balance

{\footnotesize
\bibliographystyle{IEEEtran}
\bibliography{refs_compact}

}
\newpage
\appendix

\end{document}

%% file: packages.tex
\usepackage{cite}
\usepackage{amsmath,amssymb,amsfonts}
\usepackage{algorithmic}
\usepackage{graphicx}
\usepackage{textcomp}
\usepackage{subcaption}
\usepackage{float}
\usepackage{balance}
\usepackage{grumble}
\usepackage{appendix}

\usepackage{xcolor}
\definecolor{specGray}{gray}{0.15}

\usepackage{url}
\usepackage{breakurl}
\expandafter\def\expandafter\UrlBreaks\expandafter{\UrlBreaks\do-\do_}

\graphicspath{{./figures/}}
\DeclareGraphicsExtensions{.pdf,.png}

\usepackage{listings}
\lstdefinestyle{embedded}{
	numbers=none,
	frame=none,
	xleftmargin=0cm,
	backgroundcolor=\color{Lavender},
	framesep=1pt,
	aboveskip=3pt,
	belowskip=3pt,
}
\lstdefinestyle{small}{
	basicstyle=\linespread{0.9}\footnotesize,
}

\mathchardef\hyphenmathcode=\mathcode`\-

\let\origlstlisting=\lstlisting
\let\endoriglstlisting=\endlstlisting
\renewenvironment{lstlisting}
    {\mathcode`\-=\hyphenmathcode
     \everymath{}\mathsurround=0pt\origlstlisting}
    {\endoriglstlisting}

%% file: commands.tex
\newcommand{\myparagraph}[1]{\smallskip \noindent{\bf {#1}.}}

\newcommand{\code}[1]{\mbox{\texttt{#1}}}
\newcommand{\circled}[1]{\raisebox{.5pt}{\textcircled{\raisebox{-.9pt} {#1}}}}
\newcommand{\equalDelta}[0]{$\overset{{\scriptscriptstyle\Delta}}{=}$}

\newcommand{\secref}[1]{$\S$~\ref{sec:#1}}
\newcommand{\subsecref}[1]{$\S$\ref{subsec:#1}}

\newcommand{\figref}[1]{Figure~\ref{fig:#1}}

\newcommand{\tabref}[1]{Table~\ref{tab:#1}}

\newcommand{\sys}{\textsc{T-Lease}}

\newcommand{\rdtsc}{\texttt{rdtsc}}

\newcommand\blfootnote[1]{%
  \begingroup
  \renewcommand\thefootnote{}\footnote{#1}%
  \addtocounter{footnote}{-1}%
  \endgroup
}

%% file: space-tricks.tex
\usepackage[small,compact,noindentafter]{titlesec} 

\titlespacing\section{0pt}{6pt plus 2pt minus 2pt}{2pt plus 2pt minus 2pt}
\titlespacing\subsection{0pt}{4pt plus 2pt minus 2pt}{2pt plus 1pt minus 1pt}
\titlespacing{\paragraph}{0pt}{2pt plus 0pt minus 1pt}{1.0ex}

\usepackage[inline]{enumitem}
\setlist{noitemsep,topsep=0pt,parsep=0pt,partopsep=0pt}

\usepackage[subtle]{savetrees}

\usepackage{setspace}
\usepackage[margin=5pt,font={stretch=0.9}]{caption}

%% file: 0_abstract.tex

\subsection*{Abstract}

A lease is an important primitive for building distributed protocols, and it is ubiquitously employed in distributed systems. However, the scope of the classic lease abstraction is restricted to the {\em trusted} computing infrastructure. Unfortunately, this important primitive cannot be employed in the {\em untrusted} computing infrastructure  because the trusted execution environments (TEEs) do not provide {\em a trusted time source}. In the untrusted environment, an adversary can easily manipulate the system clock to violate the correctness properties of lease-based systems.

We tackle this problem by introducing {\em trusted lease}---a lease that maintains its correctness properties even in the presence of a clock-manipulating attacker. 
To achieve these properties, we follow  a ``trust but verify'' approach for an untrusted timer,  and transform it into a trusted timing primitive by leveraging  two hardware-assisted ISA extensions (Intel TSX and SGX) available in commodity CPUs. We provide a design and implementation of trusted lease in a system called \sys{}---the first trusted lease system that achieves high security, performance, and precision.  For the application developers, \sys{} exposes an easy-to-use generic APIs that facilitate its usage to build a wide range of distributed protocols.


%% file: 1_intro.tex
\section{Introduction}
\label{sec:introduction}

Leases are one of the fundamental building blocks of distributed systems~\cite{Gray:1989:LEF:74851.74870}.  On an abstract level, a \emph{lease} is a permission to access a shared resource for a certain period of time (the {\em lease term}). The  lease is issued by an authoritative resource owner (the lease \emph{granter}) to an entity that wants to access the resource (the lease \emph{holder}). While the lease is active, the holder can freely access the resource without requiring any coordination with the granter. 
\blfootnote{This is a preprint version of a paper that was published at Middleware'20 conference.}

Due to this coordination-less scheme, leases bring a significant benefit to building distributed systems for workloads with heavy read skew by eliminating the need for repeated resource locking. Therefore, leases are ubiquitously used in the design of distributed protocols and systems, such as two-phase commit~\cite{Aguilera:2009:SNP:1629087.1629088}, locking~\cite{Yoon:2018:DLM:3183713.3196890}, consensus~\cite{DBLP:conf/wdag/Lampson96, Moraru:2014:PQL:2670979.2671001}, caching~\cite{Yu:1999:SWC:316188.316219}, leader election~\cite{Fetzer:1999:HAL:325392.325396}, failure detector~\cite{Huang:2018:CES:3291168.3291170}, databases~\cite{180324, spanner}, storage~\cite{farsite}, sharding~\cite{slicer}, and file systems~\cite{Muthitacharoen:2001:LNF:502034.502052,Kistler:1991:DOC:121132.121166,Mazieres:2001:TUF:647055.759949, DBLP:journals/cluster/HupfeldKSHCMM09, gfs}. Thanks to leases, such systems can achieve high performance and strong consistency with low overheads, while still allowing the writes to proceed. In this regard, leases are favorable compared to the traditional locking protocols for synchronization~\cite{zookeeper-atc, Burrows:2006:CLS:1298455.1298487}: with locks, the writes cannot proceed until all readers have unlocked their data.

Even though leases are widely used in distributed systems, their scope is mainly confined to the {\em trusted} computing infrastructure. However, this assumption is no longer valid with the prevalence of cloud computing: the potential risks of security violations in third-party cloud computing infrastructure have increased significantly. In the untrusted environment, an attacker can compromise the security
properties of distributed systems. Many studies show that software bugs, configuration errors, and security vulnerabilities pose a serious threat to distributed systems deployed on the untrusted computing infrastructure~\cite{Gunawi_bugs-in-the-cloud, Santos2009}. In particular, lease violations can lead to both denial of service and correctness issues (see \subsecref{t-lease}).

To mitigate the security threats in the cloud, Trusted Execution Environments (TEEs) provide an appealing way to build secure distributed systems~\cite{DBLP:journals/iacr/CostanD16,DBLP:journals/corr/abs-1907-10119,DBLP:journals/csur/PintoS19}. More specifically, Intel SGX has gained traction as a solution for bringing trust to cloud computing \cite{IBMCloud,azure_stack}. However, even with Intel SGX, the design of trusted lease is non-trivial: to enforce a lease term, TEE must have access to a trusted time source. Unfortunately, this important primitive is missing in the current versions of Intel SGX. In practice, an attacker has the capabilities to control the time sources (by setting value and frequency), the CPU frequency, by delivering interrupts and delaying messages, etc. Therefore, a strawman design for a trusted lease is bound to either be  insecure (for a design where the lease duration is measured using the Timestamp Counter) or suffer from high performance overheads (for a TPM-based clock).

To overcome these limitations, we focus on the following question---{\em how can we design a trusted lease abstraction for distributed systems}? To answer this question, we present an abstraction of \emph{trusted leases}, which we design and implement in a system called \sys{}. The trusted lease retains all properties of the classic lease~\cite{Gray:1989:LEF:74851.74870}, but it is designed to maintain its correctness properties even in the presence of a privileged attacker.  More specifically, \sys{} is the first system for trusted leases with the following design properties:
\begin{itemize}
\item {\bf Security:} It always maintains the lease correctness invariant; that is, the lease duration at the granter must be a superset of the lease duration at the holder.

\item {\bf Performance:} It imposes minimal performance overheads compared to classical leases.
\item {\bf Usability} {\em (time precision \& APIs generality)}: It provides an easy-to-use generic APIs to support  both short-termed (fine-grained time resolution) and long-termed leases for implementing a wide-range of distributed protocols.
\end{itemize}

To achieve these design goals, we apply a ``trust but verify'' approach to a high-resolution low-overhead untrusted timer, improving its security without sacrificing performance and usability. More specifically, we transform the untrusted timer into a trusted time source by leveraging ISA extensions available in commodity CPUs. This transformation is based on a simple observation: {\em the untrusted timer can only be manipulated on interrupts}; thus, \sys{} needs to detect interrupts and verify the correctness of the timer after each interrupt detection. In particular, we leverage two ISA extensions to realize our approach: Intel SGX~\cite{SGXrelease} and Intel TSX~\cite{sdm}. \sys{} relies on  Intel SGX to detect interrupts by examining the memory used for the enclave state storage during the interrupt handling. Intel TSX provides us with hardware transactional memory for rolling back an active transaction upon an interrupt delivery. A combination of these two features allows us to (a) detect interrupts before checking the lease, and (b) ensure that critical instruction sequences are executed without interrupts.

More specifically, our design builds on three core contributions: (1) {\em enclave-interval timer} allows secure and low-overhead measurement of the time intervals that the application spends inside the enclave; (2) {\em timer frequency verification} mechanism to prevent an attacker from manipulating the timer frequency, thus allowing verification of the timer correctness; (3) {\em transactional system call interface} using the hardware transactional memory to prevent the time-of-check to time-of-use (TOCTOU) vulnerability between a lease check and the corresponding system call submission.

To ensure the correctness of the \sys{} design, we formally specify the protocol and its correctness properties (safety and liveness) using TLA+~\cite{Lamport2002TLAPlus}. 
Then, we use a model checker to validate that the specification does not violate the properties.
Safety properties ensure that the semantic of a valid lease intended by the protocol is preserved, even in the presence of an attacker who manipulates the clock frequency.
Liveness properties guarantee that a lease holder eventually gets a lease using the protocol in the normal operation conditions.

We implement \sys{} as a static library, which provides an easy-to-use generic lease APIs for implementing a wide-range of distributed protocols.
In the evaluation, we study the performance, correctness, and precision properties of \sys{} using a set of microbenchmarks both in single-node and distributed system setups. We further employ \sys{} to design three real-world distributed case studies: (a) failure detector in FaRM~\cite{dragojevic2015no}, (b) Paxos Quorum Leases~\cite{Moraru:2014:PQL:2670979.2671001}, and (c) strongly consistent caching~\cite{Gray:1989:LEF:74851.74870}. The evaluation results show that \sys{} is effective at detecting timer tampering for TSC (x86 Time\-stamp Counter) and the overhead from the timer is minimal in a wide range of configurations (up to 5\% in most cases).


%% file: 2_overview.tex

\section{Overview}
\label{sec:overview}

A lease is a contract issued by a resource owner to give control to a holder over the protected resource for a certain time duration.
This duration is defined using a \emph{lease term} parameter.
A lease term might have any length, from zero to infinity.
In practice, however, the lease term is typically set to a limited amount of time. When the lease term expires, the holder usually has to renew the lease.

Typically, classical distributed systems assume trusted environments in which they rely on the system time sources, like \code{clock\_gettime} to enforce the lease term.
It provides resolution up to nanoseconds and has an extremely low overhead on modern Linux systems that use vDSO.

Compared to the classical systems, distributed systems built with TEEs assume a more privileged attacker who can affect the lease term by manipulating the system time resources.
Hence, in this paper, we introduce a novel concept of \emph{trusted leases} to tackle this challenge.

\subsection{A Case for Trusted Leases}
\label{subsec:t-lease}
The trusted lease abstraction is motivated by the necessity to secure distributed systems built using TEEs. 
TEEs provide strong confidentiality and integrity properties for the application memory but do not extend these security guarantees to the system time sources. Typically, the failure of the lease mechanism causes only denial of service. However, in the untrusted environment, a privileged attacker can manipulate the lease term as perceived by the holder or by the granter, leading to the violations of \emph{correctness}, \emph{e.g.}; system security properties. For example, leases may be used to limit the number of concurrently running enclaves (as a licensing mechanism for SGX runtime, or as a security measure, \emph{e.g.}; to limit brute-force throughput). Such use-cases can be implemented using the lease mechanism, motivating our design of trusted leases for untrusted environments.

To support such use-cases, trusted leases retain all of the properties of classic leases, but extend them with a stronger threat model, where a privileged attacker tries to subvert its correctness by influencing the runtime environment.
Therefore, in contrast to the leases for trusted environments, trusted leases cannot rely on the operating system time sources to enforce the lease term: these time sources are by definition under the control of the OS. Thus, only architectural time sources and TPMs could be used for the trusted lease implementation.

Hence, a trusted lease can be defined as a lease \emph{designed to maintain its correctness properties even in the presence of a privileged attacker}. A timer manipulation, in the worst case, results in a performance loss.

\begin{table}[t]
  \centering
  \small
  \begin{tabular}[t]{|l|l|l|c|c|}
    \hline
    Timer                                               & Type      & OS mediated& OS control & Cost \\
    \hline
    SW timer~\cite{Schwarz_2017}                        & SW        & No         & Yes& Low   \\
    TSC~\cite{sdm}                                      & Arch.     & No (SGXv2) & Yes& Low   \\
    HPET~\cite{hpet}                                    & Arch.     & Yes: MMIO  & Yes& Med.  \\
    PTP clock~\cite{Trach:2018:SSM:3185467.3185469}     & HW        & Yes: MMIO  & Yes& Med.  \\
    TPM~\cite{arthur_practical_2015}                    & HW        & Yes: OS    & No & High  \\
    \hline
  \end{tabular}
  \caption{Time sources on the x86 architecture.}
  \label{tab:time-sources}
\end{table}

\myparagraph{Threat model} We assume a powerful attacker that has full control over the OS  and can introduce arbitrary changes to the system configuration.
We focus on the attacker that manipulates clocks: changes clock value and frequency, introduces delays into application execution and message delivery, manipulates CPU frequency. Therefore, standard timers provided by the platform or the OS are untrusted \cite{DBLP:conf/uss/AnwarS19,Trach:2018:SSM:3185467.3185469}.

\tabref{time-sources} shows examples of the time sources available on the x86 architecture.
Additionally, it shows the amount of control the OS has (so, the attacker), and the overhead of the time source. 
An attacker can directly affect the time readings using a variety of mechanisms: using MMIO control registers for HPET and PTP clocks, writing to model specific registers for TSC and changing power management settings to modify the frequency of a software timer.
She also has indirect ways to affect time reading by delaying the time reading requests and pre-empting the running application.

We assume a correctly implemented CPU and ISA extensions; that is, SGX protects the confidentiality and the integrity of enclave memory and TSX aborts transactions on interrupts.
Other attacks, like buffer overflows~\cite{DBLP:conf/eurosys/KuvaiskiiOATBFF17} and side-channel attacks~\cite{216033,vanbulckq2018foreshadow,weisse2018foreshadowNG}, are out-of-scope for this paper.

\begin{figure*}[t]
  \centering
  \includegraphics{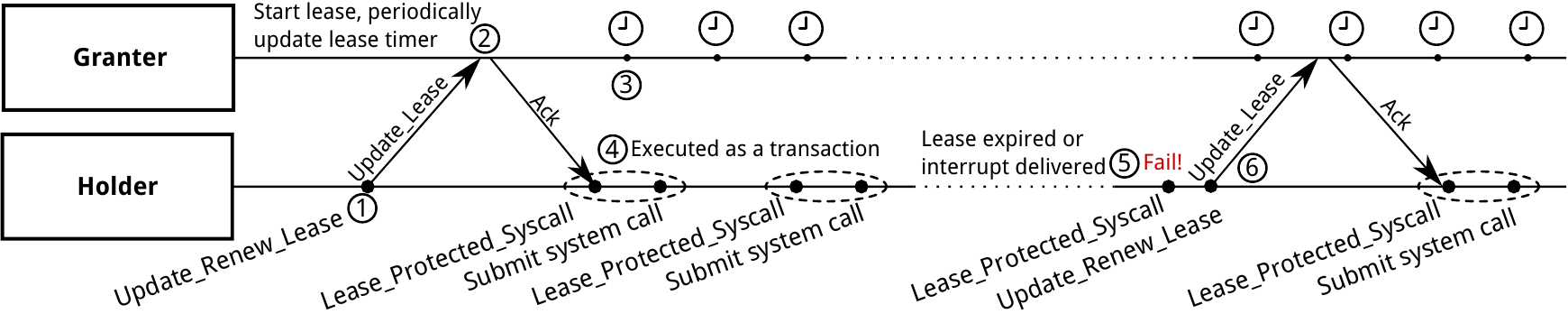}
  \caption{Basic workflow of the \sys{} protocol.}
  \label{fig:t-lease-simple}
\end{figure*}

\subsection{\sys{}: A Trusted Lease Primitive}
\label{sec:api-and-usage}

\myparagraph{\sys{} overview}
\sys{} relies on two Intel x86 ISA hardware extensions: Intel SGX\cite{SGXrelease} and Intel TSX~\cite{sdm}.
It allows building distributed lease-based applications that run inside Intel SGX enclaves and withstand attacks on the time sources by privileged adversaries.

\sys{} builds on three core abstractions:

\begin{itemize}
\item \sys{} presents an \emph{enclave-interval timer} to securely estimate the duration of the time intervals using an untrusted, OS-controllable time source. This functionality is necessary to track the lease term on the granter and the holder sides. To achieve this, the enclave-interval timer builds on the SGX architectural features for interrupt detection and uses TSC as an underlying timer.
\item \sys{} presents a \emph{timer frequency verification} mechanism for thwarting attacks that manipulate timer tick rate. To achieve this, we have designed an empirical approach that measures the duration of a sequence of attacker-uncontrollable instructions (RDRAND) and aborts execution if the result is out of architecture-determined bounds.
\item \sys{} presents a \emph{transactional system call interface} to communicate the results of computations that depend on the lease being present. So, the holder can atomically check the lease state and communicate the computed results; thus, avoiding the TOCTOU vulnerability. We achieve this by starting an Intel TSX transaction before checking the lease and committing it on a successful system call submission.
\end{itemize}

\myparagraph{\sys{} APIs} \sys{} implementation consists of a client library and a reference implementation of a lease granter. Table \ref{tab:t-lease-api} lists the core APIs that is exported by the library for the use by the granter and the holder (the service and auxiliary functions are not shown). 

The functions can be divided into the following categories: Initially, two initialization functions are used for the lease and lease protocol client: {\tt Init\_Lease} and {\tt Init\_Client}. They are called by the granter and holders to initiate working with \sys{}. Then, two core functions are used to maintain a correct \sys{} protocol state: {\tt Update\_Renew\_Lease}, called by the holder to request a lease and update the lease state, and {\tt Update\_Lease\_Client}, called primarily by the granter to update the state of the lease at its own side. {\tt Lease\_Protected\_Syscall} is a function for secure results submission: it is used by the holder to atomically check the the lease state and submit the computation results to the client. {\tt RDTSC\_AEX} is a function that is used for the low-level usage in more complex distributed protocols than the default lease protocol used in \sys{}.

\begin{table}[t]\small
  \setlength\tabcolsep{0pt}
  \centering
  \begin{tabular}{p{8.0cm}}
    \hline\\[-8pt]
    (H) {\tt Init\_Lease(Lease timeout)} \\
    \hspace*{0.5cm} Initializes a lease with configuration.\\

    \\[-8pt]
    (G, H) {\tt Init\_Client(Local Addr, Remote Addr, AES key)} \\
    \hspace*{0.5cm} Initializes client communication endpoints.\\

    \\[-8pt]
    (H) {\tt Update\_Renew\_Lease(Lease, Client)} \\
    \hspace*{0.5cm} Updates and renews the lease.\\

    \\[-8pt]
    (G, H) {\tt Update\_Lease\_Client(Lease, Current time)} \\
    \hspace*{0.5cm} Updates lease state without renewing it.\\

    \\[-8pt]
    (G, H) {\tt Lease\_Protected\_Syscall(Lease)} \\
    \hspace*{0.5cm} Enables TSX protection for system calls if lease is active.\\

    \\[-8pt]
    (G, H) {\tt T, AEX? = RDTSC\_AEX()} \\
    \hspace*{0.5cm} Reports current \rdtsc{} value and if enclave was interrupted since last call.\\

    \hline
  \end{tabular}
  \caption{APIs exported by the \sys{} library: G marks the functions used by the granter and H by the lease holder. }
  \label{tab:t-lease-api}
\end{table}

 \myparagraph{\sys{} basic workflow} 
\figref{t-lease-simple} provides the basic narration used in \sys{}.
First, since \sys{} uses TSC as a time source, which measures time in cycles, it is necessary to calculate nanoseconds-cycles conversion factors.
Then, both the holder and the granter can initialize endpoints by using the \code{Init\_Client()}. This function opens the UDP socket used for communication, and configures the cryptographic key used to secure the communication. Next, the lease holder initializes the lease: sets the requested lease term and the lease identifier. Thereafter, the holder can request the lease from the granter \circled{1}.

 The granter enters a work loop, where it first receives a message from holder, and based on the holder command activates or disables the lease \circled{2}. After serving a message from a holder, it updates the state of all active leases, by using \code{Update\_Lease\_Client} function \circled{3}. This function updates the enclave-interval timer for each lease, disabling all leases where the accumulated value is larger than the lease term.

The holder, upon receiving a lease, enters a work loop: for example, a cache server may be handling user requests. It gets the user's request, processes it, and submits the results to the user. This operation is only valid if the lease is active, hence, it needs to call \code{Lease\_Protected\_Syscall()} to check the lease state and submit the system call in a transaction manner \circled{4}. If the return value indicates that the transaction is active, the system call can be submitted. If the transaction is inactive \circled{5}, the holder needs to renew its lease using \code{Update\_Renew\_Lease} and retry \circled{6}. 
The reasons under which transactions may become inactive are explained in §~\ref{sec:untampered-time-measurement}.


%% file: 3_design.tex
\section{Design}
\label{sec:design}

In this section, we first present two strawman designs and associated design challenges to realize the trusted lease abstraction. Thereafter, we present a detailed design of \sys{}.

\subsection{Strawman Designs and Associated Challenges}
\label{sec:attacker-capab-straw-design}

\sys{} is designed to operate with Intel SGX trusted execution environments, called enclaves. As such, it may access a number of timers, presented in Table~\ref{tab:time-sources}. There is a set of trade-offs associated with each timer, that fall on the axis of the timer access cost and the control that the OS has over the timer. For example, some timers can be accessed only via the OS. For these timers, the OS can introduce arbitrary delays into message reads, so that these timers can be used only to establish the lower, but not the upper bound on the elapsed time. Other timers, like software timer and TSC with Intel SGXv2 can be accessed directly. However, the OS can use the following capabilities to subvert the timer readings.
\begin{itemize}
\item {\em Power management:} Changing the CPU frequency influences the tick rate of a software timer.
\item {\em Preempting the application or delaying messages:} This attack can be used on any OS-mediated timer.
\item {\em Modifying timer value:} Readings of TSC can be changed by writing to IA32\_TIME\_STAMP\_COUNTER or IA32\_TSC\_ADJUST model-specific registers when the enclave is preempted, readings of HPET or NIC PTP clock can be modified via MMIO writes.
\item {\em Modifying timer frequency:} On virtualized platforms, writing to TSC Multiplier and TSC Offset fields in the VM control structure changes the TSC speed~\cite{sdm}. 
\end{itemize}

All of these attacks must be thwarted by the \sys{} design, which is a non-trivial task. Consider, for example, the following two strawman solutions:

\myparagraph{TPM-based design} Consider a design where the enclave checks the lease expiration using the time read from the TPM timer. Because the OS mediates in TPM communication, such a design cannot guarantee the \emph{correctness} property. Specifically, during the lease check period, the OS can delay the TPM read beyond the lease expiration time. The holder gets the TPM read result after the lease expires, thus violating the lease invariant. Another vector for subverting system correctness is delivering an interrupt between the lease check and returning the results from the enclave. If the enclave execution resumes only after the lease expires, the lease invariant is also invalid. Besides, the TPM fails to meet the speed and accuracy goals. Reading a digitally signed TPM time takes from $50$ ms to $600$ ms depending on the selected cryptography system, i.e., hash-based or asymmetric cryptography.

\myparagraph{TSC-based design} Consider a design when  the lease is initialized, its term in seconds is converted into \rdtsc{} cycles using the CPU-specific multiplier, and requested from the granter. After the granter acknowledges the lease, the expiration point (in \rdtsc{} cycles) is calculated; as soon as this point of time elapses, the lease  becomes invalid. The granter and holder both track the lease duration.
With this design, the attacker has two prospects for subverting the security requirements. First, the OS can preempt the application, write to the Model Specific Registers to set the time inside the enclave back into the past, and then continue the application execution. Secondly, the attacker could launch the application in a VM, and use TSC Multiplier control to slow down the TSC. Next time when the enclave reads the time, it will not be able to detect the lease expiration.

\myparagraph{Design challenges} To summarize, these attack vectors present the following design challenges for \sys{}: 
\begin{enumerate}
\item How can the lease term be securely measured by the granter and the holder? ($\S$~\ref{sec:untampered-time-measurement})
\item How can the timer frequency be verified? ($\S$~\ref{sec:untampered-timer-frequency})
\item How to atomically perform the timer check and return results? ($\S$~\ref{sec:undel-reso-access})
\end{enumerate}

\subsection{\sys{} Detailed Design}
\label{sec:sys-components}

In this section, we describe the detailed design of \sys{} that addresses the aforementioned challenges.

\subsubsection{Enclave-Interval Timer}
\label{sec:untampered-time-measurement}

To help solve the first challenge, we use the following intuition: for the lease implementation, there is no need to measure the absolute time, only the relative---that is, time differences. 
To securely measure time intervals using the OS-controlled untrusted timer, which would retain the performance characteristics of the underlying clock, we need to: (a) ensure that the underlying timer was not manipulated or delayed for some period of time, and (b) precisely establish points when the manipulation could take place.

It is necessary to choose a time source. We note that all of the OS-mediated sources do not allow establishing whether the timer was not manipulated (i.e., each access is potentially manipulated), so, they cannot be used in the design of \sys{}. Thus, only software timer and TSC with SGXv2 can be used since their value or frequency can be manipulated only when the enclave is preempted. We chose to use TSC in the implementation of \sys{}, because it incurs lower performance overhead: it does not require dedication of a CPU core to a timer thread. Since in our case the resulting clock measures the duration of time intervals inside the enclave, we call this timer an {\em enclave-interval timer}.

\begin{figure}[t]
  \centering
  \includegraphics{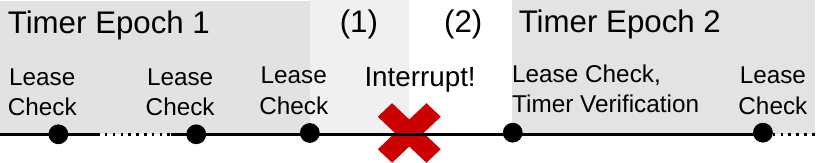}
  \caption{Enclave-interval timer operation. (1) Under-accounted time inside enclave; (2) Correctly unaccounted time outside enclave.}
  \label{fig:enclave-interval-timer}
\end{figure}

Observing the capabilities of an attacker, we come to the conclusion that the attacker needs to deliver an interrupt to tamper with system clock configuration in all cases. \sys{} uses the corollary of this fact: {\em we can safely estimate a period of time as long as the entire period is spent inside the enclave, that is as long as no interrupts happen during that period}. We call such a non-interrupted period of time an \emph{epoch}.

We detect interrupts by inspecting enclave State Save Area (SSA), a preconfigured memory region that saves the register state of enclave upon receiving an interrupt~\cite{DBLP:journals/iacr/CostanD16}. SSA has a predefined format, with fields for the registers and service data. We write $0$ (zero) into the field of the IP register, which is an invalid value for that register. Later, we can check the value of that field, and if an interrupt happened, we will detect a non-zero value there.

To estimate the duration of an active lease, \sys{} periodically reads the TSC value and checks for the interrupts using the aforementioned mechanism. If no interrupt was delivered, it adds the duration of an interval from previous such check to the current moment to the lease active time. In case there was an interrupt, the operations for the granter and holder are different. Because the lease term on granter must be a superset of the lease term on the holder, upon detecting an interrupt at the granter \sys{} can continue operation in a normal mode. This functionality is implemented in function \code{Update\_Lease\_Client}.

The holder, however, cannot do the same, because it could have been preempted for an arbitrary long period of time, and its lease on the granter could have expired in the meantime. Thus, upon each interrupt, the holder has to renew its lease from the granter. As before, the request-reply interaction should happen in the same epoch; otherwise, there is no guarantee that the packet has not been delayed. We have implemented the lease state update and communication in function \code{Update\_Renew\_Lease}.

\subsubsection{Timer Frequency Verification}
\label{sec:untampered-timer-frequency}
To solve the second challenge in which the attacker could change the TSC frequency in addition to the TSC value, \sys{} must verify the timer frequency after each interrupt.

A strawman design of the verification routine could consist of a sequence of instructions with deterministic execution time, \emph{e.g.}; noops or in-register additions. However, these actions have a significant drawback: they open a privileged attacker a possibility to tamper with the execution speed of the CPU using the power management features.

The ability of an attacker to control the power management features has far-reaching implications for the verification routine: most modern Intel CPUs have constant TSCs, that is TSC speed is independent of the CPU frequency. On the other hand, the speed of other components of the CPU does depend on the CPU frequency: by manipulating CPU frequency and \rdtsc{} speed simultaneously, an attacker can trick a simple verification routine into believing that the TSC rate is normal.

Therefore, the procedure that verifies the timer frequency must not depend on the CPU speed. By analyzing the literature\cite{rdrand:intel, rdrand:kocher} and performing experiments on multiple SGX-enabled platforms, we have discovered that the RNG module embedded into the Intel CPUs to implement RDRAND instructions is independent of the CPU frequency: entropy collection module is self-clocked at 3 GHz, and post-processing module runs unconditionally at 800 MHz. Therefore, we use a sequence of six RDRAND instructions to measure the \rdtsc{} rate. The number of RDTSC instructions to execute is a trade-off between accuracy and the verification cost.

Our measurements ($\S$~\ref{sec:microbenchmarks}) have shown that the latency variance of RDRAND is high: between 7000 and 10500 cycles. Due to an inherent variation of cost of this instruction, the attacker would still be able to modify the TSC frequency in some bounds. To increase the reliability of the rate estimation, the measurement can be repeated several times. While our verification routine depends on the microarchitectural details of the RNG, the RDRAND latency falls into these bounds on all SGX-enabled CPUs that were available to us, thus we argue that this technique is applicable in practice.

\subsubsection{Transactional System Call Interface}
\label{sec:undel-reso-access}

Finally, \sys{} has to close a window of vulnerability between the lease check and the externally-observable actions that are conditional on the lease state. In our model, we use system calls (which may involve writing to disk or sending a message over the network). We argue that this model is adequate for most of the currently used distributed systems, as the number of TEE-based systems that use kernel bypass for the communication is comparably small.

We observe that with Intel SGX, the only way for an enclave to submit computation results is via the shared memory writes. Thus, the required atomicity of the lease check and the result submission can be achieved using the hardware transactional memory: if an interrupt is delivered while the transaction is active, the underlying hardware will automatically rollback all changes made in the transaction.

\sys{} uses Intel TSX to check the lease and submit computation results in a single atomic transaction~\cite{sdm}. Intel TSX allows applications to perform arbitrary memory reads and writes in an atomic, transactional manner. TSX imposes some limitations on these transactions: the amount of writes that may happen in transaction is limited by L1 cache, some instructions inside transactions are forbidden. In case these limitations are violated, a read-write or write-write conflict is detected, or an interrupt is delivered, the transaction is rolled back with an error flag set. To limit these effects, we allow committing the transaction immediately after a system call is submitted ($\S$~\ref{sec:implementation-sys-library}).

The attacker can still delay the message or disk write after they are submitted, but this cannot violate the security properties: the messages/writes can be delayed in a distributed system even without an attack, and designing a system to tolerate these delays is out of scope of \sys{}. For synchronous and timed asynchronous system, the maximum delay must be taken into account when checking if the lease is active.


%% file: 4_implementation.tex

\section{Implementation}
\label{sec:implementation}

\subsection{Implementation of the \sys{} Library}
\label{sec:implementation-sys-library}

We implement \sys{} as a static library in 1037 lines of ANSI C, including 26 lines of inline assembly.

\myparagraph{Intel SGX framework} \sys{} relies on SCONE~\cite{scone2016} as an underlying SGX framework and to get access to the SSA region. Our work, however, is conceptually independent of SCONE and can be built on top of Graphene-SGX~\cite{Tsai:2017:GPL:3154690.3154752} and Intel SGX SDK~\cite{intel_sgx_sdk}. Other than modifying system call thread code for reducing the Intel TSX abort rate, we have added a transaction commit code in the SCONE system call handler to reduce the transaction length. 

\myparagraph{Communication} \sys{} uses UDP sockets for the communication, which is common for latency-sensitive services. All communications between the granter and the holder are encrypted with AES-GCM-256. We use Intel IPSec Multibuffer Encryption library~\cite{intel_ipsec_mb} for these cryptographic functions. \sys{} leases currently use a pre-shared AES key; in production use, we expect to use a full-fledged key management service for the key distribution.

\myparagraph{TSX-specific optimization}
When designing the TSX protection, we need to take into account the architecture of SCONE. Since SCONE uses asynchronous communication via concurrent queues between the enclave and the untrusted world, the transaction abort rate due to the read-write conflicts between the system call thread and the in-enclave thread was reaching 79\%. We have fixed this issue by adding six \code{pause} instructions into the back-off routine of the system call thread, as a trade-off between the instruction overhead and the abort rate. This has significantly reduced the abort rate of transactions---to 0.008\% on a simple system-call intensive benchmark without reducing its performance in any measurable way.

\subsection{Implementation of the \sys{} Case Studies}
\label{sec:implementation-sys-usecases}

To demonstrate how \sys{} can be used in practice, we apply it to three state-of-the-art distributed systems that rely on leases. In the following case studies, we have implemented a standalone implementation of  granters (or nodes with equivalent features); for the holder part, support for each of the use-cases was added into the client library. 

\myparagraph{Failure detector in FaRM~\cite{dragojevic2015no}} FaRM is a high-performance distributed transactional storage with high availability and strong consistency~\cite{dragojevic2015no}. FaRM uses leases as a failure detector: each worker node has to maintain a lease on a cluster manager node. When a lease expires, this signals to the lease granter that the lease holder has failed, and triggers the FaRM cluster reconfiguration. We implement the same failover protocol, recreating as many details of the original paper as possible (the lease renewal rate is set to 1/5 of lease duration, etc.). An attacker may choose to modify the time at the cluster manager node; thus, preventing the detection of the outdated leases. In this case, the cluster reconfiguration will not be updated, and the client would be directed to a node in a failed state.

\myparagraph{Paxos Quorum Leases~\cite{Moraru:2014:PQL:2670979.2671001}} Paxos Quorum Leases is a modification of the Paxos protocol that splits objects and nodes of the system into lease groups according to the frequency of accesses to each of the objects on each node~\cite{Moraru:2014:PQL:2670979.2671001}. Inside lease groups, each node has an infinite term lease to objects belonging to the group, and it can serve read accesses to these objects without consulting the majority of the nodes; thus, it significantly improves the system throughput. While the lease itself has an infinite term, to activate a lease configuration, each node must exchange a non-infinite lease with a majority of the nodes in the lease group. An attacker that manipulates the time on one or multiple machines can cause the node to assume that it has successfully established the lease with the majority of the nodes, while in practice this would not be true. By using \sys{} inside the lease activation protocol, we can ensure that the attacker cannot violate the system correctness.

\myparagraph{Consistent caching~\cite{Gray:1989:LEF:74851.74870}} Strongly consistent caching is a use-case that is commonly used in distributed systems to improve throughput and latency~\cite{Gray:1989:LEF:74851.74870}. It uses standard leases to grant a caching node access to a set of objects (files on the file system, database rows) for a lease term, during which reads from the caching node can be done without consulting an authoritative data source, and the data source will notify the caching node about any write to objects under lease. This system relies on the invariant that a lease duration at the lease holder is shorter than the lease duration at the granter. Violation of this requirement may cause stale reads or even conflicting, inconsistent results. When the strongly consistent caching is implemented with \sys{}, the manipulations of system time are detected, and the correctness of the system is ensured.


%% file: 4_modeling.tex

\section{Protocol Correctness}
\label{sec:modeling}

To validate the design of the \sys{} protocol, we provide a formal specification of the protocol and the correctness properties using TLA+ \cite{Lamport2002TLAPlus}.
Then, we use the TLC model checker \cite{Yu1999TLC} to assert the properties on a finite-instance of the specification.
	
\subsection{TLA+ Specification of the {\sys} Protocol} 
\figref{spec-overview} shows the overall structure of the specification (\texttt{\sys-Spec}) and its correctness properties.
We capture the behavior of the {\sys} protocol by modeling it as \emph{hosts} that communicate using network \emph{messages}.
Hosts include a single granter and a set of holders  defined in \texttt{LeaseHolders} set.

\myparagraph{Lease period} 
Time-based leases require the use of synchronized clocks that have a maximum drift rate of $\pm$ \texttt{Drift}.
We model the current time explicitly using \texttt{now} variable, incremented by the \texttt{Tick} action.
\texttt{LeaseTime} defines the initial lease period (lease term), taking into account drift by +\texttt{Drift} for the granter and -\texttt{Drift} for holders.
Hosts use local countdown timers (\texttt{gExpireTimer}, \texttt{lhExpireTimer}) to keep track of the lease periods.

\myparagraph{Lease grant}
Holder sends \texttt{ReqLease} message to the granter using \texttt{LHReqLeaseFresh}/\texttt{LHReqLeaseToExtend} actions by appending it to \texttt{msgs} and starts tracking the lease time.
The message includes metadata: the holder's sending timestamp \texttt{now} and \texttt{lhEpochNumber}.
The granter processes requests using \texttt{GProcessRequest} action.
If the lease is free, the granter assigns it to the holder by saving his metadata in \texttt{gLeaseGranted}.
If the lease is granted to the same holder and the request metadata is not older than saved metadata, the lease is extended, otherwise, it is rejected.
The granter responds with \texttt{Granted}/\texttt{NotGranted} message that mirrors the request metadata and the granter's sending timestamp, and starts tracking the lease time.
Once the lease expires, the granter frees the lease by setting \texttt{gLeaseGranted =\{\}}.
Using \texttt{LHReceive} action, the holder ignores the messages with smaller epoch number. 
The \texttt{Granted} message received by the holder allows him to have \texttt{validLease} state where he can safely use the resources exclusively.
Note that holder already starts the lease countdown timer (\texttt{lhExpireTimer}) at the time of sending \texttt{ReqLease}.
Therefore, if the lease expires at the holder side before receiving the \texttt{Granted}/\texttt{NotGranted} message, the validity of the lease is protected.
In this case, the holder will enter the \texttt{blocked} state and will ask for the lease again.

\input{listings/spec-overview.tex}

\myparagraph{Interrupts} 
Hosts execute inside enclaves and use transactions.
Actions \texttt{GEnclaveInterrupt} and \texttt{LHEnclaveInterrupt} allow the OS to interrupt the enclaves. 
Thus, enclaves will be in an \texttt{interrupted} state, in which their execution is temporarily halted, and causes transactions to automatically abort.
In this state, the attacker can manipulate the clock frequency by $\pm$\texttt{FreqDrift} using \texttt{AChangeFreq} action.
Actions \texttt{GEnclaveResume} and \texttt{LHEnclaveResume} resume the host's enclave execution.
Thus, holders will be in a \texttt{blocked} state since they detect that they were interrupted.
In this state, the holder is ignorant about the status of the lease due to the possible clock manipulation during enclave interrupts.
Therefore, he increments \texttt{lhEpochNumber} and sends \texttt{ReqLease} message (with the new epoch number) to the granter to renew the lease.
After resuming from the \texttt{interrupted} state, the granter can measure in a trustworthy manner only the periods spend inside the enclave, therefore, he decrements the lease period only inside the enclave.
Thus, the lease might expire at the holder side earlier than it expires at the granter side.

\subsection{Correctness Properties}

We validate the protocol design by validating safety and liveness properties.
Validating safety property ensures the protocol will not be in any state that violates the correctness of the lease even when an attacker exists. 
Validating liveness properties ensures that in the absence of an attacker, the protocol allows holders to make progress by exclusively acquiring the lease.

\myparagraph{Safety}
\figref{spec-overview} (lines \ref{line:s_s}-\ref{line:e_s}) provides the formal definition of the lease validity property \texttt{ValidLease}.
The lease is considered valid, if for all time units in which the holder has the lease (\texttt{validLease}) that did not yet expire (\texttt{lhExpireTimer[h]$\neq$ 0}), the granter also has the record that the lease is granted to the same holder (\texttt{r.lh = h}).

\myparagraph{Liveness}
Ensuring the protocol makes progress in the absence of an attacker and using bounded execution time frame (\texttt{MaxNow}) requires multiple \emph{assumptions} to hold.
Specifically, we assume bounded message delivery delay (\texttt{MsgDeliveryMaxDelay}), that the OS will not starve the enclave execution (\texttt{InterruptedMaxPeriod}), and that the enclave will not be interrupted for at least \texttt{NotInterruptedMinPeriod}.

\figref{spec-overview} specifies two liveness properties in lines \ref{line:s_p1}-\ref{line:e_p1} and \ref{line:s_p2}-\ref{line:e_p2}.
\texttt{HolderAsksForLeaseGranterGrantsLease} specifies that if a holder sends a message to the granter, the granter should eventually (\texttt{MaxNow}) grant him the lease if it is not granted to anyone else (\texttt{gLeaseGranted $\neq$ \{\}}).
Additionally, the message must be sent before \texttt{MaxNow} with enough time to reach the granter (\texttt{(MaxNow - now) >= MsgDeliveryMaxDelay}) and the granter should be in a processing state (\texttt{gEpochTimer > (MaxNow - now)}). 

\texttt{GranterGrantsLeaseHolderHasValidLease} specifies that if the granter granted the lease to a holder, the holder should eventually (\texttt{MaxNow}) have a \texttt{validLease} state. 
It requires assumptions similar to the first property,  additionally, it assumes that the holder should not be interrupted between sending the request and receiving the response (\texttt{lhEpochNumber[h] = r.epochNumber}).

\subsection{Validation using the TLC Model Checker}

To prove that the specification satisfies the correctness properties, the TLC model checker exhaustively enumerates all possible system states and verifies that none of the states violates the given properties.
In the absence of an attacker, the \texttt{\sys-Spec} and a finite-instance of the specification where all hosts use equal \texttt{leaseTime}, TLC reports it has not found any violation of the safety and liveness properties.
However, if the attacker changes the clock frequency to maximum $\pm 50\%$ (see \secref{microbenchmarks} (b)),
TLC shows that if the granter uses \texttt{leaseTime} that is three times larger than the holder's (\texttt{gExpireTimer} = $3$ * \texttt{lhExpireTimer}), then, the lease validity is preserved.



%% file: listings/spec-overview.tex
\begin{figure}[t!]

\begin{lstlisting}[escapeinside={(*}{*)}, mathescape=true]
(*\textcolor{specGray}{\textbf{CONSTANTS}}*) LeaseHolders, Drift, LeaseTime, 
 FreqDrift, MaxNow, NotInterruptedMinPeriod,
 MsgDeliveryMaxDelay, InterruptedMaxPeriod, ...
(*\textcolor{specGray}{\textbf{Vars}}*) (*\equalDelta*) $\langle$ now, msgs, lhState, gState, lhEpochNumber, (*\label{line:s_vars}*)
  lhExpireTimer, gExpireTimer, gLeaseGranted, 
  lhEpochTimer, gEpochTimer, lhFrequency ... $\rangle$ (*\label{line:e_vars}*)
(*\textcolor{specGray}{\textbf{Messages}}*) (*\equalDelta*) [msgType: {"ReqLease"}, h:LeaseHolders,(*\label{line:s_messages}*)
  epochNum: Nat, timeStamp: Nat] $\cup$  msgType:
  {"Granted", "NotGranted"}, h: LeaseHolders,
  epochNum: Nat, timeStamp: Nat, sendTimeStamp: Nat](*\label{line:e_messages}*) 
(*\textcolor{specGray}{\textbf{TypeOK}}*) (*\equalDelta*) $\wedge$ lhState $\in$ [LeaseHolders $\mapsto$(*\label{line:s_typeok}*)
  {"created", "pending", "validLease", "blocked", 
   "interrupted"}]
  $\wedge$ gState $\in$ {"insideEnclave", "interrupted"}
  $\wedge$ gLeaseGranted $\subseteq$ [lh: LeaseHolders,
    timeStamp: Nat, epochNumber: Nat] $\cup$ {} 
  $\wedge$ msgs  $\subseteq$ Messages $\cup$ {}   $\wedge$ ...(*\label{line:e_typeok}*)
(*\textcolor{specGray}{\textbf{Init}}*) (*\equalDelta*)   $\wedge$ now = 0 $\wedge$ msgs = {}(*\label{line:s_init}*)
  $\wedge$ lhState = [h $\in$ LeaseHolders $\mapsto$ "created"]
  $\wedge$ gState = "insideEnclave"   
  $\wedge$ lhEpochNumber = [h $\in$ LeaseHolders $\mapsto$ 1]
  $\wedge$ lhExpireTimer = [h $\in$ LeaseHolders $\mapsto$ $\infty$]
  $\wedge$ gExpireTimer = $\infty$   $\wedge$ gLeaseGranted = {} $\wedge$ ...(*\label{line:e_init}*)
(*\textcolor{specGray}{\textbf{Next}}*) (*\equalDelta*)  $\vee$ Tick $\vee$ ($\exists$ h $\in$ LeaseHolders:(*\label{line:s_next}*)
  $\vee$ LHReqLeaseFresh(h)$\vee$ LHReqLeaseToExtend(h) 
  $\vee$ LHReceive(h) $\vee$ LHEnclaveInterrupt(h) 
  $\vee$ GProcessRequest(h) $\vee$ AChangeFreq(h) 
  $\vee$ LHEnclaveResume(h)) $\vee$ GLeaseExpires 
  $\vee$ GEnclaveInterrupt $\vee$ GEnclaveResume (*\label{line:e_next}*)
(*\textcolor{specGray}{\textbf{Fairness}}*) (*\equalDelta*) $\wedge$ SF$_{now}$(Tick)  $\wedge$ $\forall$ h $\in$ LeaseHolders: (*\label{line:s_fairness}*)
 SF$_{Vars}$( LHReceive(h) $\vee$ LHEnclaveResume(h) $\vee$ ...)
  $\wedge$ $\forall$ h $\in$ LeaseHolders: WF$_{Vars}$( GLeaseExpires $\vee$ ...)(*\label{line:e_fairness}*)
(*\textcolor{specGray}{\textbf{\sys-Spec}}*) (*\equalDelta*) Init $\wedge$  $\square$[Next]$_{Vars}$ $\wedge$ Fairness (*\label{line:spec}*)
(*\textcolor{lightgray}{--------------------------------------------------------------------------}*)
(*\textcolor{specGray}{\textbf{ValidLease}}*) (*\equalDelta*) (*\label{line:s_s}*) $\forall$ h $\in$ LeaseHolders: (
 (lhState[h] = "validLease" $\wedge$ lhExpireTimer[h] $\neq$ 0)
  $\Rightarrow$  ( $\exists$ r $\in$ gLeaseGranted: r.lh = h ))(*\label{line:e_s}*)
(*\textcolor{specGray}{\textbf{HolderAsksForLeaseGranterGrantsLease}}*) (*\equalDelta*)(*\label{line:s_p1}*)
($\exists$ m $\in$ msgs: m.msgType $\in${"ReqLease"} $\wedge$ (MaxNow-now)
$\geq$ MsgDeliveryMaxDelay $\wedge$ gEpochTimer > (MaxNow-now))
 $\leadsto$  ( gLeaseGranted $\neq$ {} )(*\label{line:e_p1}*)
(*\textcolor{specGray}{\textbf{GranterGrantsLeaseHolderHasValidLease}}*) (*\equalDelta*)(*\label{line:s_p2}*)
( $\exists$ h $\in$ LeaseHolders:  $\exists$ r $\in$ gLeaseGranted: r.lh = h
 $\wedge$  gExpireTimer = LeaseTime  $\wedge$ (MaxNow - now)$\geq$
  MsgDeliveryMaxDelay  $\wedge$ (lhEpochNumber[h]= 
  r.epochNumber) $\wedge$ lhEpochTimer[h]>(MaxNow - now)) 
  $\leadsto$ ($\exists$ h $\in$ LeaseHolders: lhState[h]={"validLease"})(*\label{line:e_p2}*)
\end{lstlisting} 

\caption{The structure of {\sys} specification and properties in TLA+ .}
\label{fig:spec-overview}
\end{figure}

%% file: 5_evaluation.tex

\section{Evaluation}
\label{sec:evaluation}

\begin{figure*}[t]
  \begin{minipage}[t]{0.33\textwidth}
    \centering
    \includegraphics[width=\columnwidth]{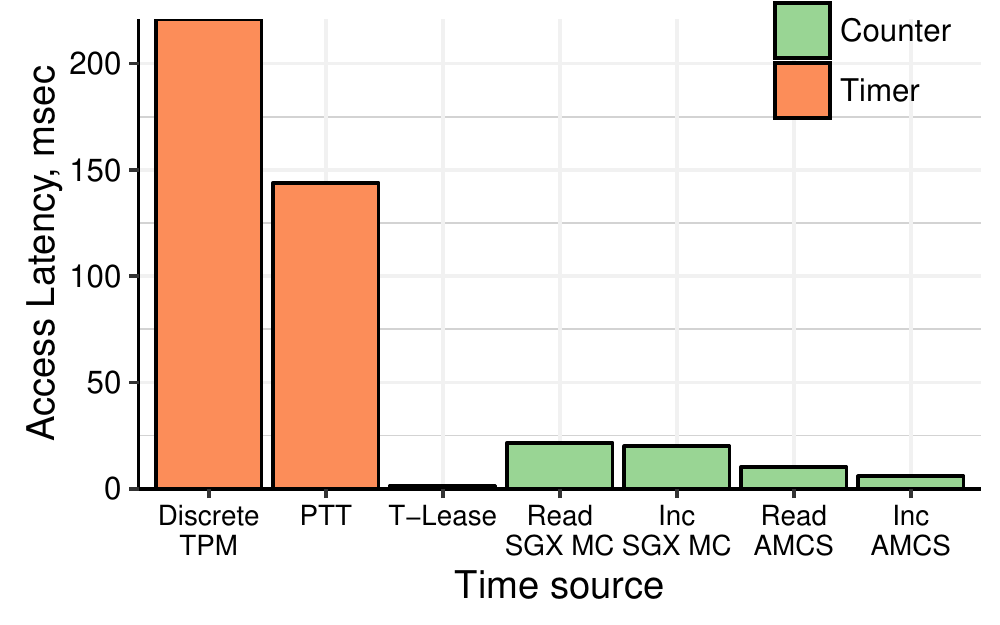}
    \caption{Access latency of trustworthy clocks and timers.}
    \label{fig:clock-latency}
  \end{minipage}\hspace{0.05cm}
  \begin{minipage}[t]{0.33\textwidth}
    \centering
    \includegraphics[width=\columnwidth]{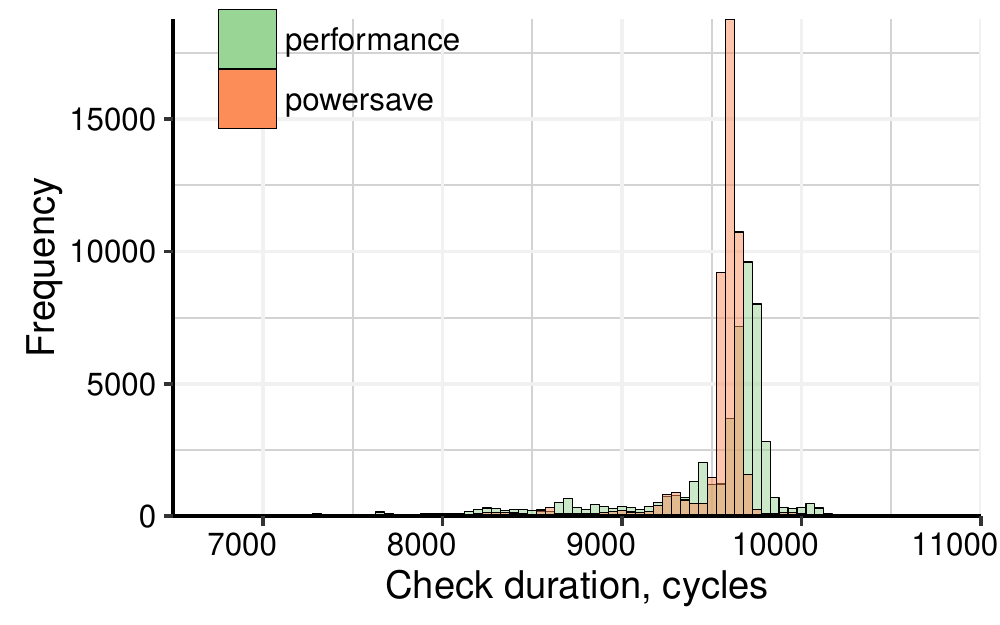}
    \caption{Latency of TSC timer check using 6 \code{rdrand} instructions.}
    \label{fig:check-latency}
  \end{minipage}\hspace{0.05cm}
  \begin{minipage}[t]{0.33\textwidth}
    \centering
    \includegraphics[width=\columnwidth]{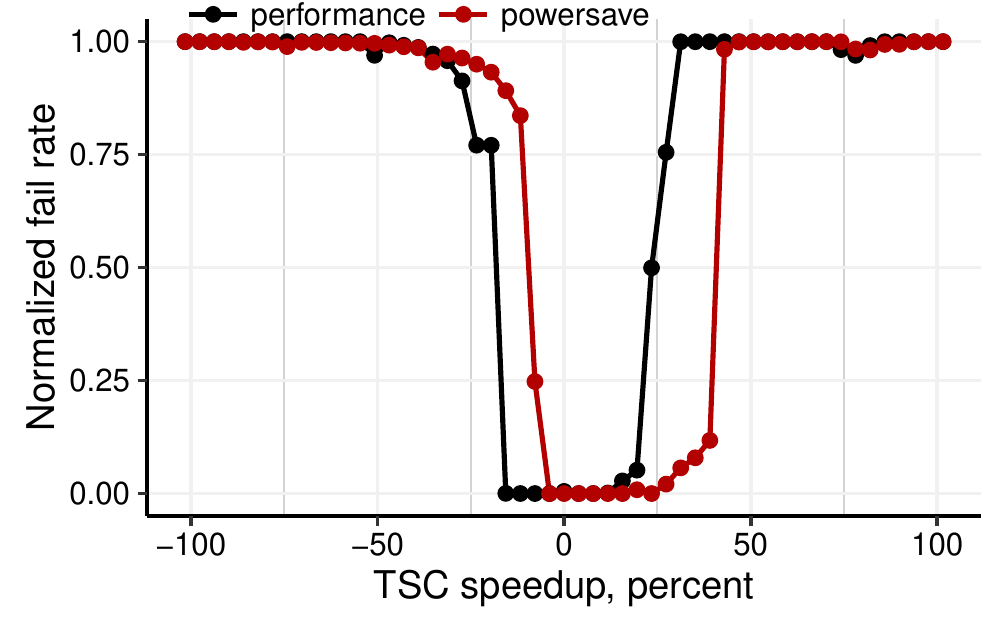}
    \caption{Probability of detecting the TSC rate manipulation.}
    \label{fig:check-probability}
  \end{minipage}
\end{figure*}

\myparagraph{Testbed} We use two types of machines. \emph{SGXv2 NUC} is an SGXv2-capable Intel Pentium Silver NUC (Gemini Lake) operating at 1.5 GHz with 16 KB L1, 128 KB L2, and 4 MB L3 caches, and 32 GB of RAM\@. \emph{SGXv1 server} is a Dell PowerEdge R330 server with an SGXv1-capable Intel Xeon E3-1270 v5 CPU (Skylake), with 32 KB L1, 256 KB L2, and 8 MB L3 caches, 64 GiB RAM, and a discrete Infineon 9665 TPM 2.0. We use the SGXv1 server in distributed experiments.
Both machines are connected to a 1~Gb/s switched network.

\myparagraph{Methodology} As system interrupts have a major influence on the functioning of the \sys{} timer, we reconfigured the system to reduce their frequency.
We changed the kernel configuration to the lowest possible timer interrupt frequency (100 Hz), enabled the dynamic ticks kernel mechanism, and steered all device interrupts to core 0.
Because of the high performance impact of interrupts on SGX enclaves, these changes are generally beneficial to SGX-based systems~\cite{Costan2016}.

\begin{figure*}[t]
  \begin{minipage}[t]{0.33\textwidth}
    \centering
    \includegraphics[width=\columnwidth]{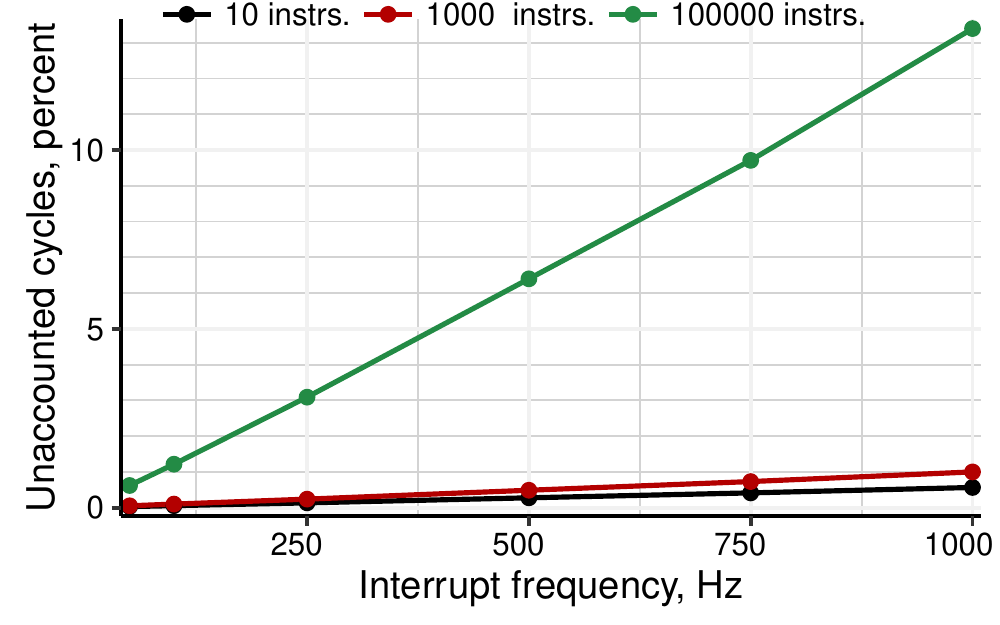}
    \caption{Cycle under-accounting for changing instr. counts b/w lease updates.}
  \label{fig:check-latency}
  \end{minipage}\hspace{0.05cm}
  \begin{minipage}[t]{0.33\textwidth}
    \centering
    \includegraphics[width=\columnwidth]{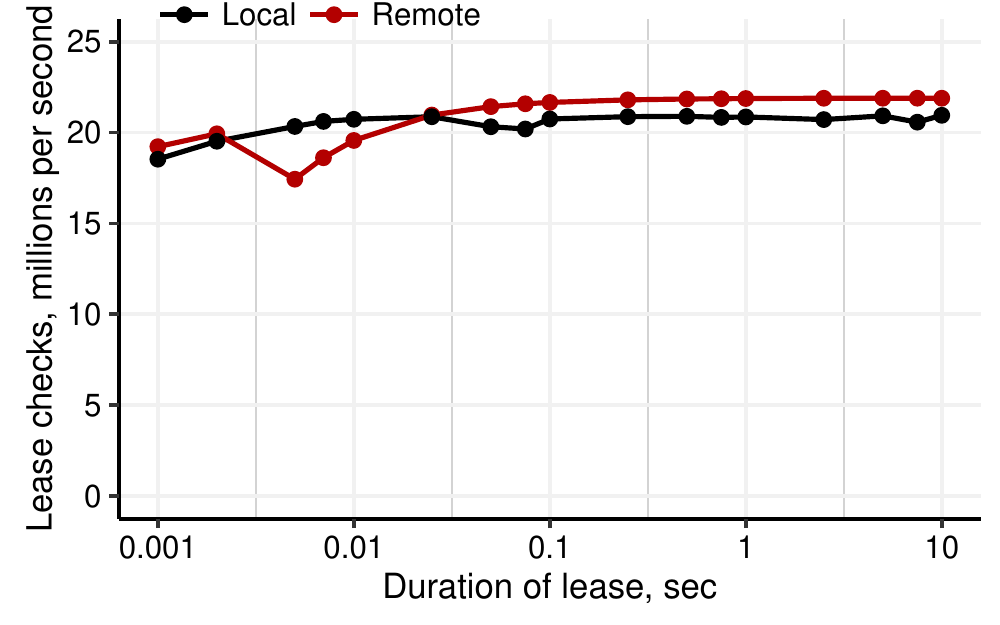}
    \caption{Frequency with which client can check \sys{} for expiration.}
    \label{fig:check-frequency}
  \end{minipage}\hspace{0.05cm}
  \begin{minipage}[t]{0.33\textwidth}
    \centering
    \includegraphics[width=\columnwidth]{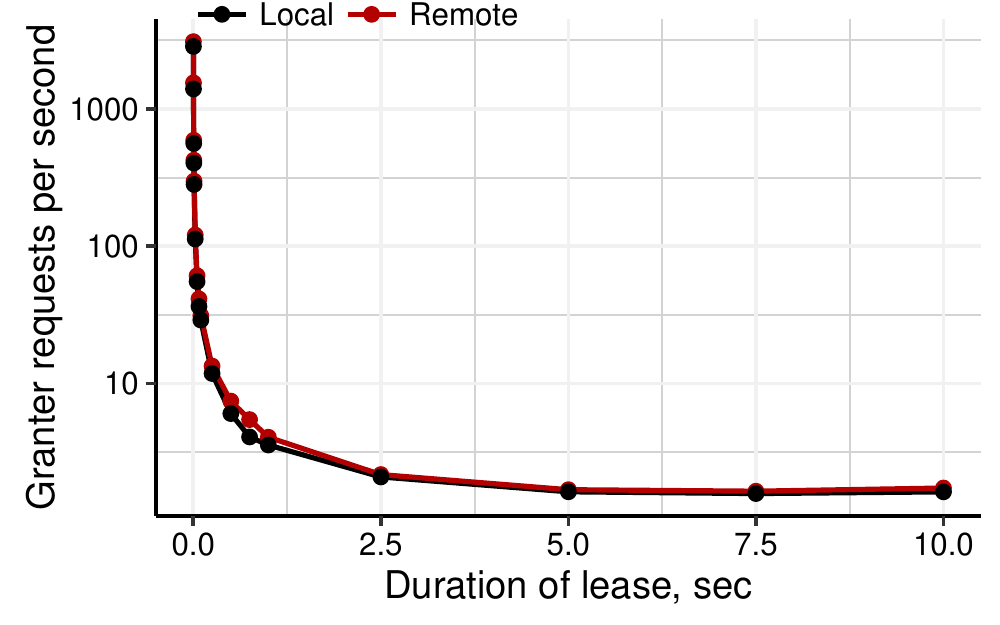}
    \caption{Frequency of network requests from the holder to the granter.}
    \label{fig:request-frequency}
  \end{minipage}
\end{figure*}

\subsection{Single-node Setup}
\label{sec:microbenchmarks}
We first evaluate (a) performance, (b) correctness, and (c) precision properties of \sys{} in a single-node setup.

\myparagraph{(a) Performance: Access latency} We begin by measuring the access latency of available secure clocks and timers (\figref{clock-latency}). Without interrupts, the access latency for \sys{} is \textasciitilde30~ns. When interrupts are delivered, the minimum cost is approximately 10k cycles (20.3~\textmu{}s) and the full cost of using a timer will depend on the interrupt recovery actions. This latency is significantly lower than the latency of TPM-based timers (220~ms for discrete TPM, and 145~ms for Intel PTT).

To put the results into a perspective, we also show the access latencies of several \emph{counter} implementations, which may also be used in a distributed system for message ordering and conflict resolution. SGX MC is the Intel SGX SDK monotonic counter implementation, which is built using Intel ME. AMCS is the network service that exposes the Intel SGX MC over the network, with on-disk caching for counter values (both counters are discussed in \secref{related}). \sys{} performs favorably to these systems as well.

\myparagraph{(a) Performance: Epoch duration} To apply \sys{} in practice, the epoch duration must be large enough for both communication with the granter and performing useful work. We have evaluated an average duration of the lease on the configured to reduce the interrupt frequency. We have discovered that in-enclave thread running with normal priority have an average epoch duration of 15~ms; when a thread is running with a real-time priority, it achieved an average duration of 650~ms. Thus, in further experiments we have configured the enclave threads to run with real-time priority. In general, this change is also beneficial for the SGX enclaves since frequent enclave exits significantly reduce enclave performance~\cite{Costan2016}.

\myparagraph{(b) Correctness: TSC rate estimator} Next, we evaluate the operation of \sys{} \rdtsc{} rate estimator (\secref{untampered-timer-frequency}). Since an attacker can break the operation of a naive rate estimator by changing the CPU frequency, we perform the experiment at two extreme CPU frequency values (800MHz and 1.5GHz on SGXv2 NUC). If the estimator performs correctly for both of these values, \sys{} will operate correctly when the frequency is set to any of the intermediate values. We change the CPU frequency between the experiments by setting the system frequency scaling governor to one of the following values: \code{performance} for 1.5GHz, and \code{powersave} for 800MHz.

Because the latency of \code{rdrand} and \rdtsc{} is not fully deterministic, there is noise in the measurements. To illustrate it, we measure the latency distribution of our estimator (\figref{check-latency}). The check duration is largely independent of the CPU frequency, which matches the documentation~\cite{rdrand:intel}. Thus, we can use these operations to implement a TSC rate estimator. Execution of the estimator should take between 7.5k and 10.5k cycles.

However, a range of 3,000 cycles is still large enough to allow an attacker to manipulate the TSC rate. We evaluate the bounds in which the attacker can successfully manipulate the tick rate without being detected (Figure~\ref{fig:check-probability}). In this experiment, we measure the probability with which \sys{} estimator will declare \rdtsc{} manipulation depending on the change in the tick rate. We can see that the attacker can slow down the timer by approximately 40\% or speed it up by 45\% with a high success probability. To protect against this manipulation, we need to either make the lease $\frac{1 + 0.45}{1 - 0.45} = 2.64$ times shorter at the lease holder, or increase its length at the granter by the same factor. We increased the lease duration at the granter by a factor of 2, which is a trade-off between attack detection probability and lease term extension, and which we use in all further experiments.

\myparagraph{(c) Precision: Under-accounting due to interrupts} For leases to operate correctly, the granter's lease term must be longer than the holder's. However, this can cause a precision issue even if there is no attack: interrupt causes under-accounting of time by the interval starting from the last time update or lease check; this extends the lease by the lost interval (Region (1) in Figure~\ref{fig:enclave-interval-timer}). We measure the amount of time that is lost depending on the interrupt rate on the core, and the number of instructions between the interval timer updates. The results are in Figure \ref{fig:check-latency}. We see that if the lease is checked in a tight loop, the loss of precision is minimal even at the extreme interrupt frequencies. If the granter is doing a significant work between the timer updates, it can incur up to 14\% precision loss. Yet, we note that the normal interrupt frequency on Linux is 250Hz for desktop and 100Hz for the server configuration, so even in the case of significant work between the timer updates, the lease extension should stay within 5\% of the nominal lease term.

\subsection{Distributed Setup}
\label{sec:evaluation-trusted-leases}

\begin{figure*}[t]
  \begin{minipage}[t]{0.33\textwidth}
    \centering
    \includegraphics[width=\columnwidth]{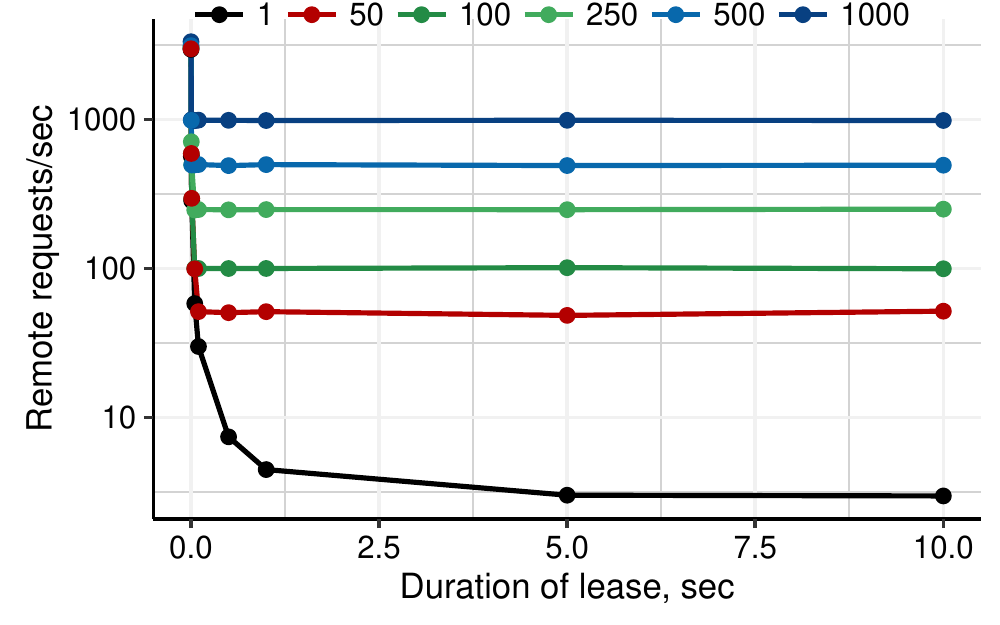}
    \caption{Influence of interrupt rate (HZ) on the request rate.}
    \label{fig:interrupt-rate-influence}
  \end{minipage}\hspace{0.05cm}
  \begin{minipage}[t]{0.33\textwidth}
    \centering
    \includegraphics[width=\columnwidth]{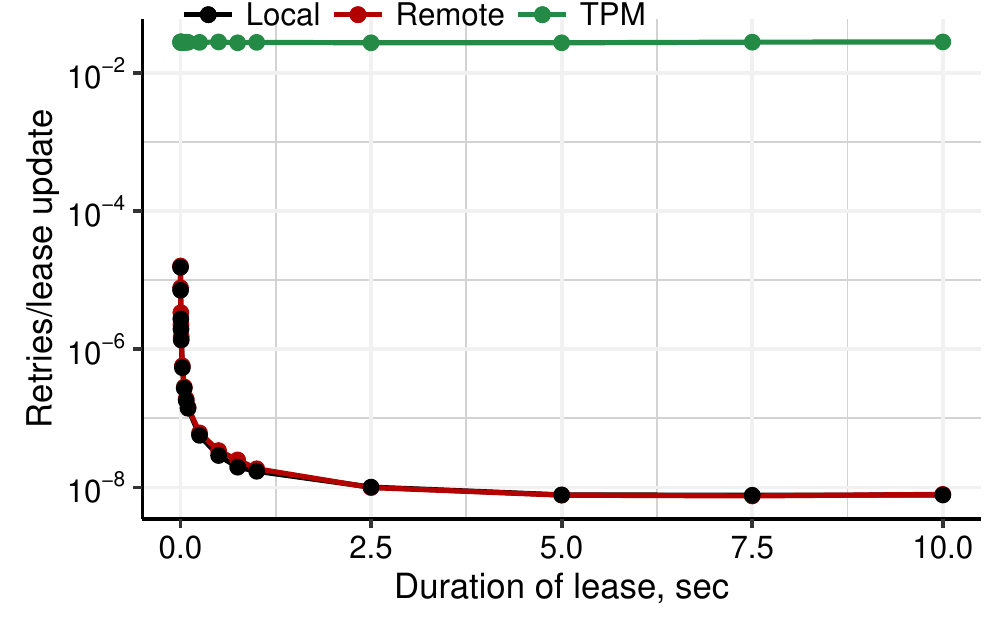}
    \caption{Frequency of retries due to interrupt delivery during lease renewal.}
    \label{fig:retries}
  \end{minipage}\hspace{0.05cm}
  \begin{minipage}[t]{0.33\textwidth}
    \centering
    \includegraphics[width=\columnwidth]{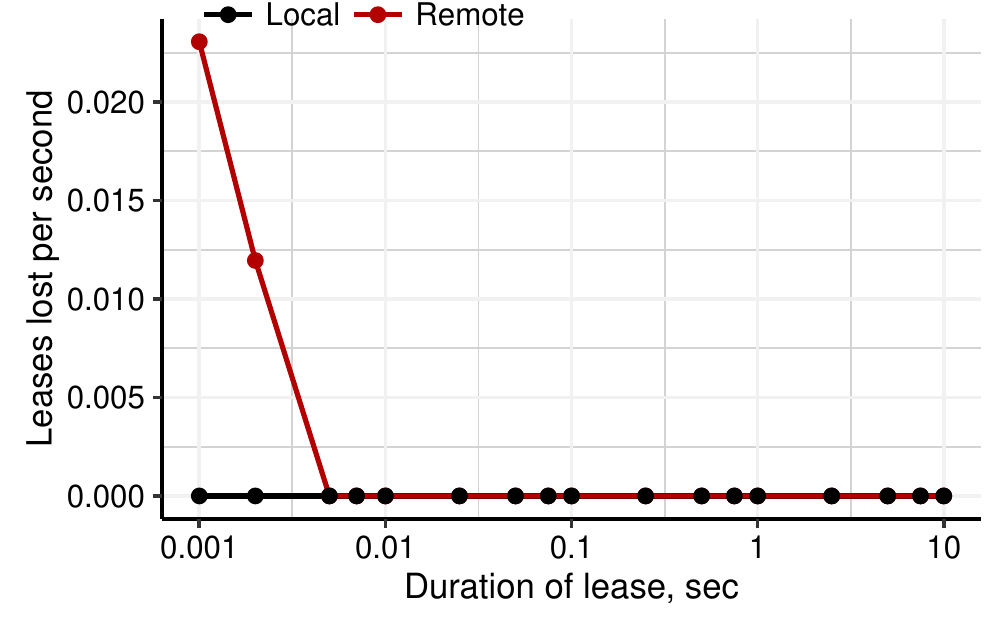}
    \caption{Number of lost leases per second for  the local \& remote setups.}
    \label{fig:loss-count}
  \end{minipage}
\end{figure*}

We next present an evaluation of trusted leases in a distributed setup with various system configurations.
We run the experiments in both local (granter and holder on the same machine) and remote (different machines) setups to estimate the network latency impact.

\myparagraph{Lease check frequency} First, we measure the frequency with which the lease holder can check the lease expiration (\figref{check-frequency}). The measurement shows whether \sys{} can become a bottleneck if the checks are located on a hot path. As we can see, the frequency has a minor dependency on the lease duration: The longer the leases are, the less frequent lease requests become and the less time is spent waiting on granter responses. In the remote setup, network delays also decrease the check rate, but only up to the lease duration of 12.5 ms, after which the effect becomes negligible. Notably, with longer leases, the remote setup has higher check rates compared to the local setup because the local does not have free cores for system tasks, thus causing higher interrupt rates.

\myparagraph{Frequency of remote requests} We also measure the rate lease extension requests (\figref{request-frequency}). They happen when either a lease expires at the holder, or when an interrupt ends the holder's epoch. We can see that the message rate is driven mostly by the lease duration for very short leases; with longer leases, interrupts maintain the minimum message rate. In this experiment, we have not discovered any difference between the local and remote setups.

\myparagraph{Impact of interrupts} In \sys{}, the holder must request a new lease after every interrupt, causing an increased lease request rate if the system issues frequent interrupts. We evaluate this property in \figref{interrupt-rate-influence}.
While typical Linux systems are configured to have a timer interrupt rate between 100 and 250 Hz, devices such as disks can generate interrupts at a much higher rate. In this measurement, we estimate the message rate in a local setup that results from different interrupt rates, from 1 to 1000 Hz. They cover a wide range of usage scenarios, from a mostly idle server (1 Hz) to a server overloaded with interrupts (1000 Hz). For longer leases, the interrupt rate determines the communication rate in all of these cases; for the short leases, the communication rate is driven by the lease expiration. For high interrupt rates, the system may experience a high message load (2000 messages/s for response and reply).

\myparagraph{Lease acquisition retries} If an interrupt is delivered before the response from the granter arrives, the holder sends one more request. We measure the average number of retries due to such interrupts, normalized by the total number of lease checks (\figref{retries}). With \sys{} timer, the number of retries is negligible (below $10^{-6}$ for leases longer than 100~ms). When instead a TPM is used as a trusted time source, the average retry rate is 0.28 per lease renewal. It proves our previous claim that TPM cannot be used to efficiently implement the lease service as-is.

\myparagraph{Lost leases} Finally, we evaluate \sys{} performance as a failure detector (\figref{loss-count}). To this end, we measure the rate of lease expiration in spite of lease holder being active (i.e., false positives). The lease expires when the holder is descheduled for a long time, or if the packets with lease renewal messages are delayed or lost in the network. \sys{} performs without lost leases in the local communication case. However, in case of communication over network, leases with terms shorter than 5~ms exhibit a false positive rate of around 1 lost lease every 2~s. In practice, network delays are taken into account when choosing a lease duration~\cite{Gray:1989:LEF:74851.74870}, which allows minimizing the lease loss.

\subsection{Case Studies}
\label{sec:use-cases}

\begin{figure*}[t]
  \begin{minipage}[t]{0.33\textwidth}
    \centering
    \includegraphics[width=\columnwidth]{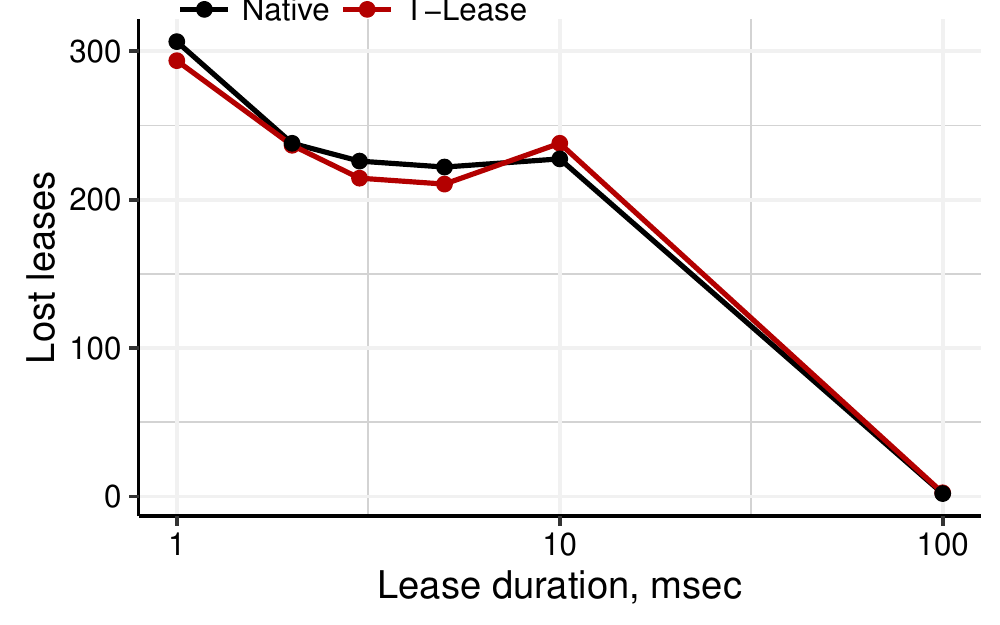}
    \caption{Number of lost leases for the FaRM failover protocol.}
    \label{fig:farm-lost-leases}
  \end{minipage}\hspace{0.05cm}
  \begin{minipage}[t]{0.33\textwidth}
    \centering
    \includegraphics[width=\columnwidth]{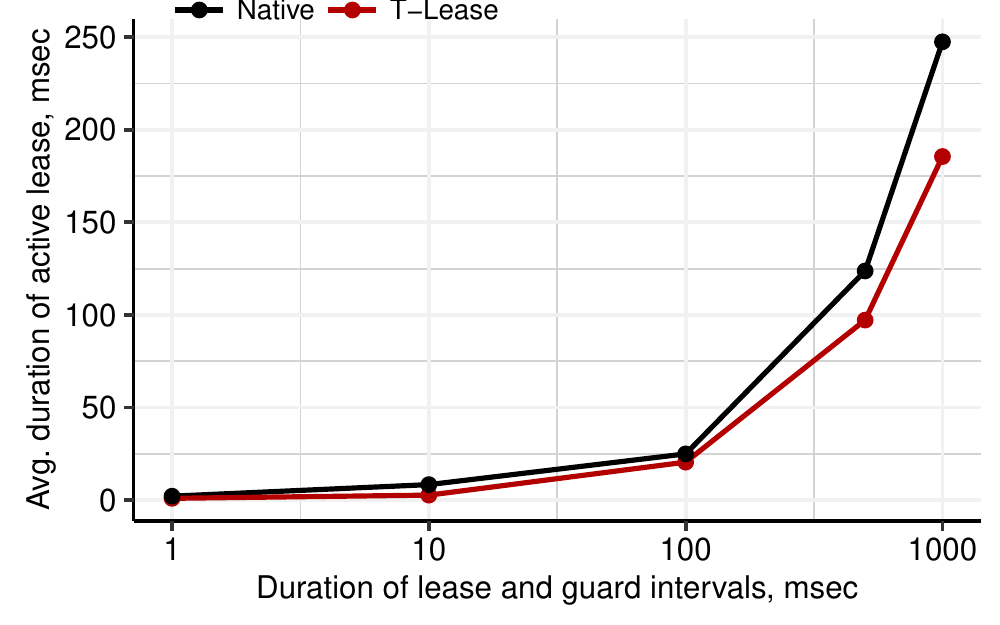}
    \caption{Duration of timer interval with active lease for the PQL case study.}
    \label{fig:usecase-pql-retries}
  \end{minipage}\hspace{0.05cm}
  \begin{minipage}[t]{0.33\textwidth}
    \centering
    \includegraphics[width=\columnwidth]{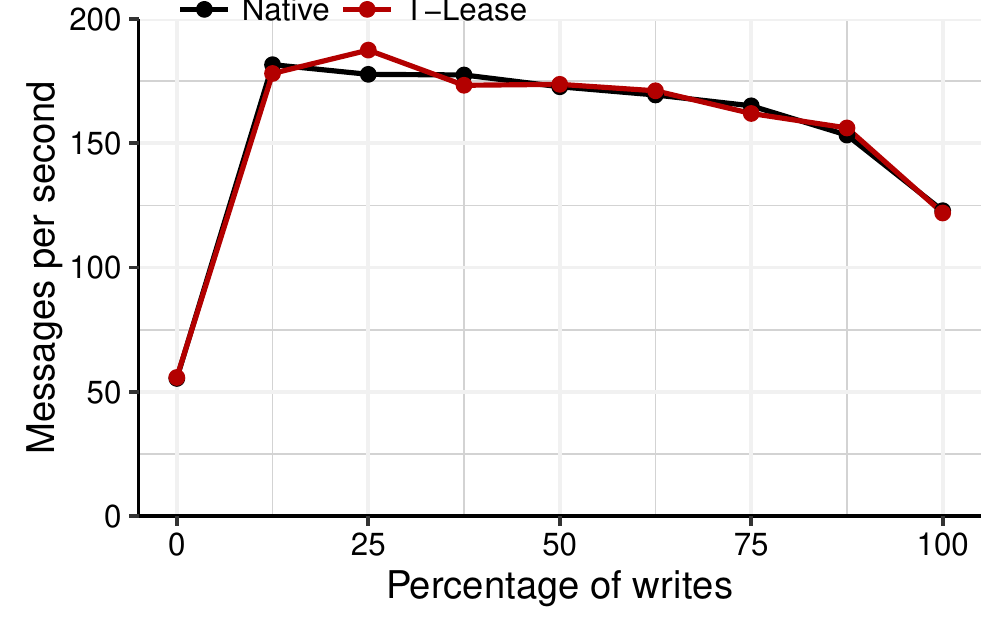}
    \caption{Average message rate for the strongly consistent caching case study.}
    \label{fig:mesi-messages}
  \end{minipage}
\end{figure*}

To showcase \sys{} in realistic conditions, we applied it to time-critical components of several distributed systems as outlined in \secref{implementation-sys-usecases}. In this section, we compare the performance of the use-cases as implemented with \sys{} to the implementations using standard leases.

\myparagraph{Failure detector in FaRM~\cite{dragojevic2015no}} Following the original FaRM paper, we measure the amount of lease expirations (i.e. due to message delays or thread inactivity) over 10 minutes (Figure~\ref{fig:farm-lost-leases}). FaRM uses unreliable datagrams over RDMA for transport, while \sys{} relies on UDP/IP and Ethernet, thus rendering the direct number comparison meaningless; nevertheless, the comparison of \sys{} with its version without interrupt detection allows us to determine the overhead of \sys{}. Unlike the original FaRM, \sys{} achieves operation without lost leases only at 100~ms lease duration. On the other hand, the lease loss rate is the same in native and the \sys{} cases.

\myparagraph{Paxos Quorum Leases (PQL)~\cite{Moraru:2014:PQL:2670979.2671001}} We implement the protocol in two variants: \sys{} and native. Because interrupt delivery may cause additional lease requests and affect performance, we measure the average duration of a lease depending on guard and lease interval, set to the same value (\figref{usecase-pql-retries}). As the lease interval increases, the active lease duration also increases, albeit in smaller steps. The native variant has slightly longer durations of active intervals, as the interrupts cause lease invalidation. This makes active interval for long leases close to 247 msec, while it is 185~ms in case of \sys{}.

\myparagraph{Strongly consistent caching service~\cite{Gray:1989:LEF:74851.74870}} Our strongly consistent caching service implementation follows that of the original lease paper. We configure the system such that the granter is running on SGXv1 machine outside of the enclave, and two cache nodes are running on SGXv2 machine. One of the cache nodes is acquiring only read leases, while the second node acquiring read or write leases with a controlled probability. We measure the average number of messages per second over a period of 5 minutes (\figref{mesi-messages}). As the number of writes in the system becomes non-zero, the messaging rate in the system increases from 55 to \textasciitilde{}175 msg/s. This is caused by frequent changes between reading and writing lease states, causing frequent lease invalidation and re-establishment. As the write share reaches 100\%, the messaging rate decreases, as the writer submits requests faster than the reader reacquires the lease.


%% file: 7_related.tex
\section{Related Work}
\label{sec:related}

\myparagraph{Trusted hardware for distributed systems}
The pioneering systems that used commodity trusted hardware for securing distributed protocols were TrInc~\cite{DBLP:conf/nsdi/LevinDLM09} and Assayer~\cite{ParnoThesisBook}. TrInc uses hardware-provided trusted counters to protect against equivocation attacks. Unlike \sys{}, it uses counters, not timers, and proposes non-standard hardware extensions. Assayer relies on the standard TPM hardware, which it leverages to convey end-host information to the network in a trustworthy and efficient manner. Both of these systems are not designed to secure lease-based protocols. Pasture~\cite{DBLP:conf/osdi/KotlaRRSW12} is a system for providing secure offline data access, allowing a remote party to audit the data access log. As cryptographic keys are used to access the data, Pasture uses TPM for key and log management. Memoir \cite{DBLP:conf/sp/ParnoLDMM11} uses a conceptually similar state continuity technique, which relies on TPM in its operation.

\myparagraph{Monotonic counters}
Another important problem that shares design space with \sys{} is protecting storage systems from rollback attacks. This is typically accomplished using monotonic counters. Intel SGX SDK contains implementations of monotonic counters  using Intel Management Engine~\cite{intel_sgx_sdk}. However, the performance of Intel SGX SDK monotonic counters is insufficient for applications. Therefore, systems that use monotonic counters either cache the counter value on the disk, or exploit workload properties to update the counter asynchronously as proposed in Speicher~\cite{DBLP:conf/fast/BailleuTBFHV19}. An alternative approach that promises to overcome the  performance and security limitations of NVMEM-based monotonic counters is building a distributed consensus-based trusted counters such as proposed in ROTE~\cite{DBLP:conf/uss/MateticAKDSGJC17}.

\myparagraph{Trustworthy timers}
Aurora~\cite{liang2018aurora} addresses the trustworthiness requirement by building design on the System Management Interrupts (SMI) to access the timer through a trustworthy environment (System Management Mode).
However, it has high costs because an application is preempted when the SMM software is running. Several systems \cite{Chen:2017:DPS:3052973.3053007, 216033, Schwarz_2017} require high-precision low-latency clock to measure cache access time required to mitigate side-channel attacks.
To that end, they employ a timer thread that increments a memory location in a tight loop. 
However, they are prone to false positives as the CPU frequency is often changing due to the power saving features.
Lastly, S-FaaS~\cite{alder2018sfaas} is a trustworthy serverless platform build using Intel SGX. For trusted CPU time accounting, it also measures the duration of time intervals but accomplishes this using a timer thread running on the sibling hyperthread, which has more overhead than \sys{}.

\myparagraph{Low-overhead timers for Intel SGX}
To address the performance and accuracy issues, non-enclave software can use the TSC or HPET for fast and high-resolution time reads~\cite{dpdk:gsg}.
In the context of SGX enclaves, SafeBricks~\cite{211281} proposes to use timestamps set on packets by the NIC as the time source to avoid slowdown due to high latency of TSC in SGXv1.
Similarly, ShieldBox~\cite{Trach:2018:SSM:3185467.3185469} uses an on-NIC PTP clock. However, both of these clock implementations are untrusted.


%% file: 6_discussion.tex

\section{Discussion}
\label{sec:discussion}
We next discuss some design extensions and choices.

\myparagraph{Compiler support}
We currently implement \sys{} as a library. Therefore, the developer has to spend additional effort to instrument timer invocations in the application to use our library calls.
However, we believe, it is possible to avoid this effort by automatically transforming the calls to \code{rdtsc} and system time into our library calls.
In the future, we plan to implement this transformation as a LLVM compiler pass to transparently benefit existing applications.

\myparagraph{Upper measurement bound}
As discussed before, \sys{} estimates only the lower bound for long-term measurements and our design cannot provide any guarantees regarding the upper bound.
It might become an issue if the upper bound is required for the algorithm correctness.
For example, in our lease system, this restriction forces us to renew the lease after every interrupt, which might lead to an excessive number of messages in the system and an overload on the granter side when there are many holders with frequent system interrupts.

\myparagraph{Hardware extensions}
The design of \sys{} could be further simplified and optimized using a simple hardware extension to the TSC functionality.
For example, if TSC would include a separate read-only register incremented every time the TSC value is modified, our system would be able to provide precise timing for much longer intervals.
This extension would reduce the rate of messages in the leasing system and improve the throughput of the timestamp service.
Also, if the hardware would expose the TSC frequency to the user-space applications, the timer verification mechanism would not be necessary.


%% file: 8_conclusion.tex
\section{Conclusion}
\label{sec:conclusion}

In this paper, we introduced a concept of a trusted lease, a variant of the classical lease that maintains its correctness properties in the presence of a privileged attacker. We designed and implemented \sys---a trusted lease system for Intel SGX enclaves. \sys{} exposes an easy-to-use interface that allows system designers to implement a wide range of trusted distributed lease-based protocols for the untrusted computing infrastructure. To achieve our design goals, \sys{}  relies on three core contributions: (a) enclave-interval timer for secure measurement of time intervals which are free from manipulations,  (b) a timer frequency verification routine that detects manipulations of TSC speed, and (c) transactional syscall interface for atomic lease state check and resource access. \sys{} implements these abstractions using Intel SGX and Intel TSX architecture extensions. We formally validate the correctness properties of the \sys{} protocol. Our evaluation with a wide range of state-of-the-art distributed protocols shows that in most cases \sys{}  adds up to 5\% overhead, thus allowing its practical utilization in modern distributed systems.

\section*{Acknowledgments}

We thank the anonymous reviewers for their helpful comments. This project was funded by the DFG Grant 389792660 as part of TRR 248, and by the EU H2020 Programme under the LEGaTO Project (legato-project.eu), grant agreement No. 780681.
